\documentclass{article} 
\usepackage{iclr2023_conference,times}

\usepackage{amsmath,amsfonts,bm}

\def\eqref#1{equation~\ref{#1}}

\def\1{\bm{1}}

\DeclareMathAlphabet{\mathsfit}{\encodingdefault}{\sfdefault}{m}{sl}
\SetMathAlphabet{\mathsfit}{bold}{\encodingdefault}{\sfdefault}{bx}{n}

\usepackage{hyperref}
\usepackage{url}
\usepackage[utf8]{inputenc} 
\usepackage[T1]{fontenc}    
\usepackage{wrapfig,lipsum,booktabs, graphicx}       
\usepackage{amsfonts, amsmath}       
\usepackage{nicefrac}       
\usepackage{microtype}      
\usepackage{xcolor}         
\usepackage{multirow}       
\usepackage{gnuplottex}
\usepackage{wrapfig}
\usepackage{hyperref}       
\usepackage{cleveref}
\usepackage{natbib}

\newtheorem{theorem}{Theorem}[section]
\newtheorem{definition}{Definition}[section]

\usepackage{mathtools}

\newcommand{\wh}{\widehat}

\newcommand{\G}{\mathcal{G}}

\newcommand{\I}{\mathcal{I}}
\newcommand{\X}{\mathcal{X}}
\newcommand{\V}{\mathcal{V}}
\newcommand{\EE}{\mathcal{E}}
\newcommand{\W}{\mathcal{W}}
\newcommand{\M}{\mathbf{M}}
\newcommand{\RR}{\mathbb{R}}

\title{Multiparameter Persistent Homology for Molecular Property Prediction}

\author{Andac Demir, Bulent Kiziltan \\
AI \& Computational Sciences (AICS), Novartis Institutes for BioMedical Research (NIBR)\\
Cambridge, MA, 02139, USA \\
\texttt{\{andac.demir,bulent.kiziltan\}@novartis.com} \\
}

\iclrfinalcopy 
\begin{document}

\maketitle

\begin{abstract}
In this study, we present a novel molecular fingerprint generation method based on multiparameter persistent homology. This approach reveals the latent structures and relationships within molecular geometry, and detects topological features that exhibit persistence across multiple scales along multiple parameters, such as atomic mass, partial charge, and bond type, and can be further enhanced by incorporating additional parameters like ionization energy, electron affinity, chirality and orbital hybridization. The proposed fingerprinting method provides fresh perspectives on molecular structure that are not easily discernible from single-parameter or single-scale analysis. Besides, in comparison with traditional graph neural networks, multiparameter persistent homology has the advantage of providing a more comprehensive and interpretable characterization of the topology of the molecular data. We have established theoretical stability guarantees for multiparameter persistent homology, and have conducted extensive experiments on the Lipophilicity, FreeSolv, and ESOL datasets to demonstrate its effectiveness in predicting molecular properties.
\end{abstract}

\section{Introduction}
Drug discovery process is the early phase of the pharmaceutical R\&D pipeline that takes $10-15$ years end-to-end and costs in excess of $\sim2$ billion US dollars~\citep{berdigaliyev2020overview}. Early phases of drug discovery process begin with identifying possible drug targets (gene/protein) and their role in the disease progression. Target identification is followed by high-throughput screening (HTS) experiments (via either structure- or ligand-based virtual screening methods) to identify the drug candidates within large compound libraries that affect the target in the desired way, also called as ``hits'' or ``leads''. In the final phases of the R\&D pipeline, drug candidates have to pass a series of rigorous controlled tests in clinical trials to be considered for regulatory approval by FDA. Overall, drug discovery is a complex and error-prone process. Specifically, in oncology only $4\%$ of therapies entering phase $1$ clinical trials ultimately gains provisional approval from FDA~\citep{mullard2016parsing}  due to experimental disproof of the promised therapeutic efficacy or unforeseen side effects. Another challenging problem which arises in this domain is that the known chemical space including public databases and proprietary data owned by private organizations probably contains on the order $100$ million molecules, 
while the chemical space might contain as many as $10^{60}$ compounds obeying Lipinski’s rule-of-five for oral bioavailability~\citep{reymond2012exploring}.

Molecular property prediction has received substantial interest in recent years to accelerate the drug discovery process~\citep{stokes2020deep} and predict the 3D structure of proteins~\citep{jumper2021highly}, with models showing potential to solve contemporary problems in materials science~\citep{schmidt2019recent} and quantum chemistry~\citep{dral2020quantum}. Most early studies utilized canonical compound representations such as SMILES~\citep{zheng2019identifying} Morgan fingerprints~\citep{zhang2019lightgbm} or eigenspectrum of Coulomb matrices~\citep{montavon2012learning} as low level molecular descriptors. More recently, variants of Graph Neural Networks (GNN) such as GIN~\citep{xu2018powerful, peng2020enhanced}, GAT~\citep{velickovic2017graph, wang2021drug} and MPNN~\citep{yang2019analyzing} exhibited state-of-the-art performance for molecular representation learning across several compound datasets on bioactivity and molecular property prediction. Despite their success, utilization of GNNs for molecular property prediction suffers from $2$ severe limitations:
\begin{enumerate}
    \item Neighborhood information is typically aggregated by permutation invariant, but non-injective operations such as an average, sum or max. This leads to an oversmoothing problem, where node embeddings converge to similar values and the information-to-noise ratio of the message received by the nodes decreases~\citep{chen2020measuring, ogawa2021adaptive}. 
    \item GNNs operating on the chemical structure of molecules hold the underlying assumption that atoms are connected together via chemical electronic bonds (ionic and covalent bonds). In fact, spatial arrangement of atoms in a molecule is also affected by the ubiquitous but subtle van der Waals forces between atoms. Although the van der Waals forces are comparatively weaker forces than ionic and covalent bonds, they play a fundamental role in specifying the molecular geometry, also known as conformation. A number of studies have shown that using $3D$ molecular conformers significantly  improves the accuracy of molecular property prediction~\citep{schutt2017schnet, gasteiger2020directional, liu2021spherical, gasteiger2021gemnet}. However, generating several low energy stable $3D$ molecular conformers is computationally infeasible for large scale applications~\citep{xu2021learning, shi2021learning, ganea2021geomol}. Some alternative methods extract more $3D$ information by using bond lengths~\citep{chen2019utilizing}, bond angles~\citep{gasteiger2020directional} or torsion angles~\citep{gasteiger2021gemnet} as edge features.          
\end{enumerate}

Our main objective is to investigate whether topological data analysis (TDA) tools, in particular persistent homology (PH), can overcome the limitations of GNNs for molecular representation learning. In this paper, we propose utilizing multiparameter persistent homology to produce novel topological fingerprints of molecules and evaluate their performance as well as suitability for property prediction on benchmark datasets: Lipophilicity, FreeSolv and ESOL. Persistent homology analyzes the topological features of a graph at multiple scales, represented by a filtration of the graph. A filtration is a sequence of nested subspaces of the data, where each subspace includes the previous ones and adds new elements. The topological features are then calculated for each subspace in the filtration, and the ones that persist across multiple scales are considered to be significant. 

We use the Vietoris-Rips (VR) graph filtration to construct a sequence of simplical complexes of a molecular graph. VR graph filtration is a simple and effective way, as it only requires specifying a distance threshold and does not require any prior knowledge of the topology of the data besides the shortest path distances of all node pairs. In this approach, the number of bonds between two vertices (atoms) is used as a measure of the distance between them. For example, if there is a direct edge (a bond) between two vertices, the distance between them is considered to be 1. If there is no direct edge between two vertices, the distance between them is considered to be infinity. In the VR graph filtration, a molecular graph is constructed by connecting vertices that are within a specified distance threshold from each other. This graph represents the first subspace in the filtration. The next subspace is then obtained by increasing the distance threshold and adding new connections to the graph. This process is repeated to obtain a sequence of nested graphs, which form the filtration. This results in a sequence of graphs that capture the topological features of the original graph at different scales.

This paper extends our previous work, namely \textbf{ToDD}, for ligand based virtual screening, which predicts the binding affinity between a target protein and a small molecule with no prior information of the $3D$ structure of the target protein~\citep{demir2022todd}. 

\medskip
{\bf The key contributions of this paper are: } 
\begin{enumerate}
    \item We develop a novel method to generate molecular fingerprints using persistent homology, which reveals hidden structures and relationships in the molecular geometry and uncovers topological features that persist across multiple scales. The most important advantage of persistent homology over GNNs is that it provides a more comprehensive and interpretable characterization of the topology of the data. 
    \item We extend traditional persistent homology by analyzing topological features that persist across multiple scales along multiple parameters (atomic mass, partial charge and bond type). Additionally, these parameters can be augmented using periodic
    properties such as ionization energy and electron affinity as well as molecular information such as chirality, orbital hybridization, number
    of Hydrogen bonds or number of conjugated bonds at the cost of computational complexity. Our fingerprinting method based on multiparameter persistent homology  reveals new insights into the molecular structure that are not easily apparent from a single parameter or scale analysis. 
    \item We establish theoretical guarantees for the stability of compound fingerprints extracted by multiparameter persistent homology.
    \item We perform extensive experiments on Lipophilicity, FreeSolv and ESOL datasets to demonstrate the effectiveness of the proposed compound fingerprints for molecular property prediction.
\end{enumerate}

\section{Related Work}
There are $2$ main approaches to predict the physical and chemical properties of molecules: 1) Applying well-known models like support vector machines or gradient boosting regression trees to expert-engineered descriptors or molecular fingerprints and 2) optimizing the model architecture of GNNs.

In the first approach, the models are applied to molecular fingerprints, such as the Dragon descriptors~\citep{mauri2006dragon} or Morgan (ECFP) fingerprints~\citep{rogers2010extended}. One direction of improvement in this approach is to use domain expertise and augment the feature representation of nodes (atoms) with more chemical information. Additionally, some studies have used explicit $3D$ atomic coordinates to further improve performance~\citep{schutt2017schnet, kondor2018covariant, faber2017machine, feinberg2018potentialnet}.

In the second approach, the focus is on optimizing the model architecture and improving neighborhood aggregation. One such model is the Graph Convolutional Neural Network (GCN), which learns the compound's feature representation by the convolution operations performed in the spectral domain of the compound's $2D$ graph, which is obtained by transforming the graph into a set of eigenvectors and eigenvalues. GCN has been shown to be flexible and capable of capturing complex relationships~\citep{wu2018moleculenet}. Another one, the Message Passing Neural Network (MPNN) framework presented in \citep{gilmer2017neural} operates by passing messages between nodes in the graph, updating the representations of the nodes and edges in the process. The message passing process is performed multiple times, allowing the model to build up a more complex and informative representation of the graph.

Recently, conformer generation has become an important step in property prediction because the physical and chemical properties of a molecule depend on its $3D$ structure. For example, the solubility of a molecule depends on its ability to dissolve in a solvent, and this ability can be influenced by the shape and orientation of the molecule. Similarly, the reactivity of a molecule can be influenced by the orientation of its functional groups, which can affect its ability to participate in chemical reactions. Therefore, conformer generation allows us to make a more accurate prediction of molecular properties~\citep{axelrod2020molecular}. This is especially important for molecules that have flexible or non-rigid structures, as they can adopt different conformations in different environments. $3D$ Infomax~\citep{stark20223d} pre-trains a $2D$ network  by maximizing the mutual information (MI) between its representation
of a molecular graph and a $3D$ representation produced from the molecules’ conformers. The weights of $2D$ network are then fine-tuned to predict properties. 

\begin{figure}
\centering
\includegraphics[width=\textwidth]{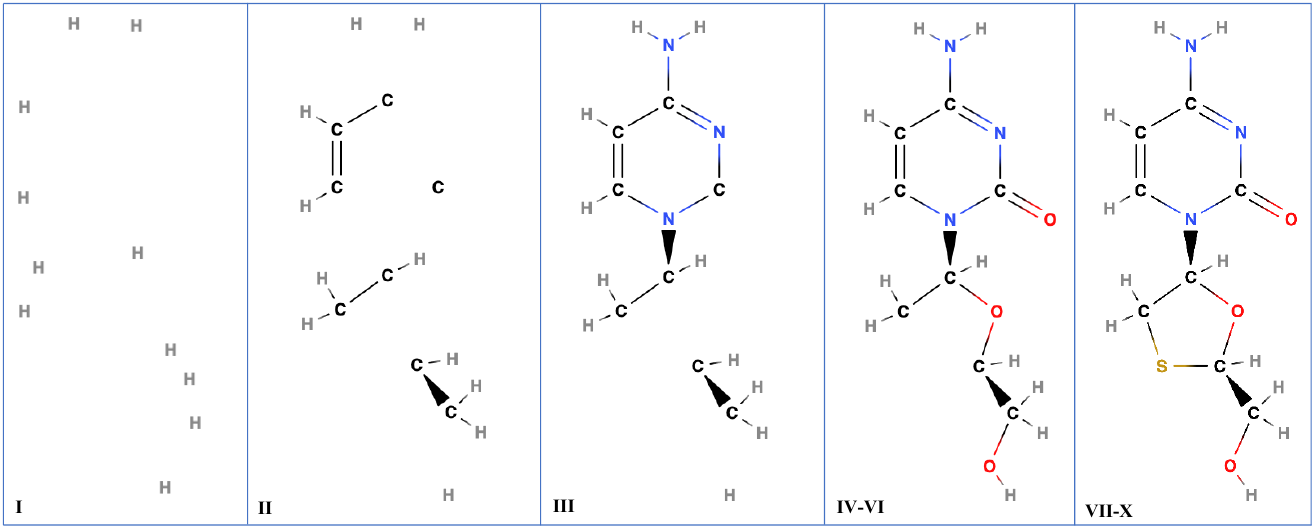}
\caption{\textbf{Graph Decomposition} \scriptsize masks some vertices based on the values of the vertices' parameters and considers only the remaining vertices and edges. In this dataset, $10$ unique atoms (H, C, N, O, F, P, S, Cl, Br, I) were present. Hence, there are $10$ subgraphs created (some subgraphs can be identical in this process). The original compound shown in the last column is lamivudine, an antiviral used to treat hepatitis B. Vertices with the highest atomic mass are masked first and they are added with respect to ascending values of atomic mass. Atoms are coded by their color: Gray=Hydrogen, Black=Carbon, Blue=Nitrogen, Red=Oxygen, Yellow=Sulfur. Figure~\ref{fig:Graph_Filtrations} is representative of the graph evolution along the y-axis in the initial column of Figure~\ref{fig:VR_Filtrations}. Our framework establishes the Vietoris-Rips complexes for every subgraph as demonstrated along the rows of Figure~\ref{fig:VR_Filtrations}. Subsequently, it computes the rank of the homology groups of dimension 0, 1 for each sequence of simplicial complexes. We repeat the same process using parameters: partial charge (with respect to the ascending order of decile groups in partial charge histogram) and bond types (initially, vertices that constitute a ring structure are added, followed by the addition of vertices linked by triple bonds, double bonds, and ultimately, single bonds). This process of masking vertices and creating subgraphs provides valuable insights into the relationships between the atom/bond properties and the topological features of the dataset.} 
\label{fig:Graph_Filtrations}
\end{figure}

\section{Persistent Homology} \label{sec:SPH}
Here, we introduce single parameter persistent homology. In essence, the process of persistent homology consists of three steps. The first step is \textit{graph decomposition} by masking vertices based on either ascending or descending order of their values, which breaks down a graph into many smaller subgraphs. In the second step, the \textit{persistent homology} machinery tracks the changes (on each subgraph separately) in topological features such as birth and death times as they occur in a sequence of simplicial complexes. Finally, in the \textit{vectorization} step, these records can be transformed into a vector that can be utilized in machine learning models.


\noindent {\bf Graph Decomposition:} For a given unweighted graph (compound) $\G=(\V,\EE)$ with $\V = \{v_1, \dots, v_m\}$ the set of nodes (atoms) and $\EE =\{e_{rs}\}$ the set of edges (bonds), we decompose $\G$ into many subgraphs using a function $f:\mathcal{V}\to\RR$ with threshold sets $\I=\{\alpha_i\}_{i=1}^m$, where $\alpha_1=\min_{v \in \V} f(v)<\alpha_2<\ldots<\alpha_m=\max_{v \in \V} f(v)$. For $\alpha_i\in \I$, let $\V_i=\{v_r\in\V\mid f(v_r)\leq \alpha_i\}$ (sublevel sets for $f$). This defines a hierarchy $\V_1 \subset\V_2\subset  \dots \subset \V_m=\V$ among the nodes with respect to the function $f$ and yields a nested sequence of subgraphs as illustrated in Figure~\ref{fig:Graph_Filtrations}. For molecular machine learning applications, this filtering function $f$ can be atomic mass, partial charge, bond type, electron affinity, ionization energy or another important function representing chemical properties of the atoms or bonds. One can also use the natural graph induced functions like node degree, betweenness, etc. 

\noindent {\bf Vietoris-Rips (VR) Filtration:} This section outlines our technique for constructing VR simplicial complexes for $2D$ multipersistence, which can be generalized to $3D$ or higher dimensions, though such an extension falls beyond the scope of this paper.

\begin{figure}
\centering
\includegraphics[width=0.75\textwidth]{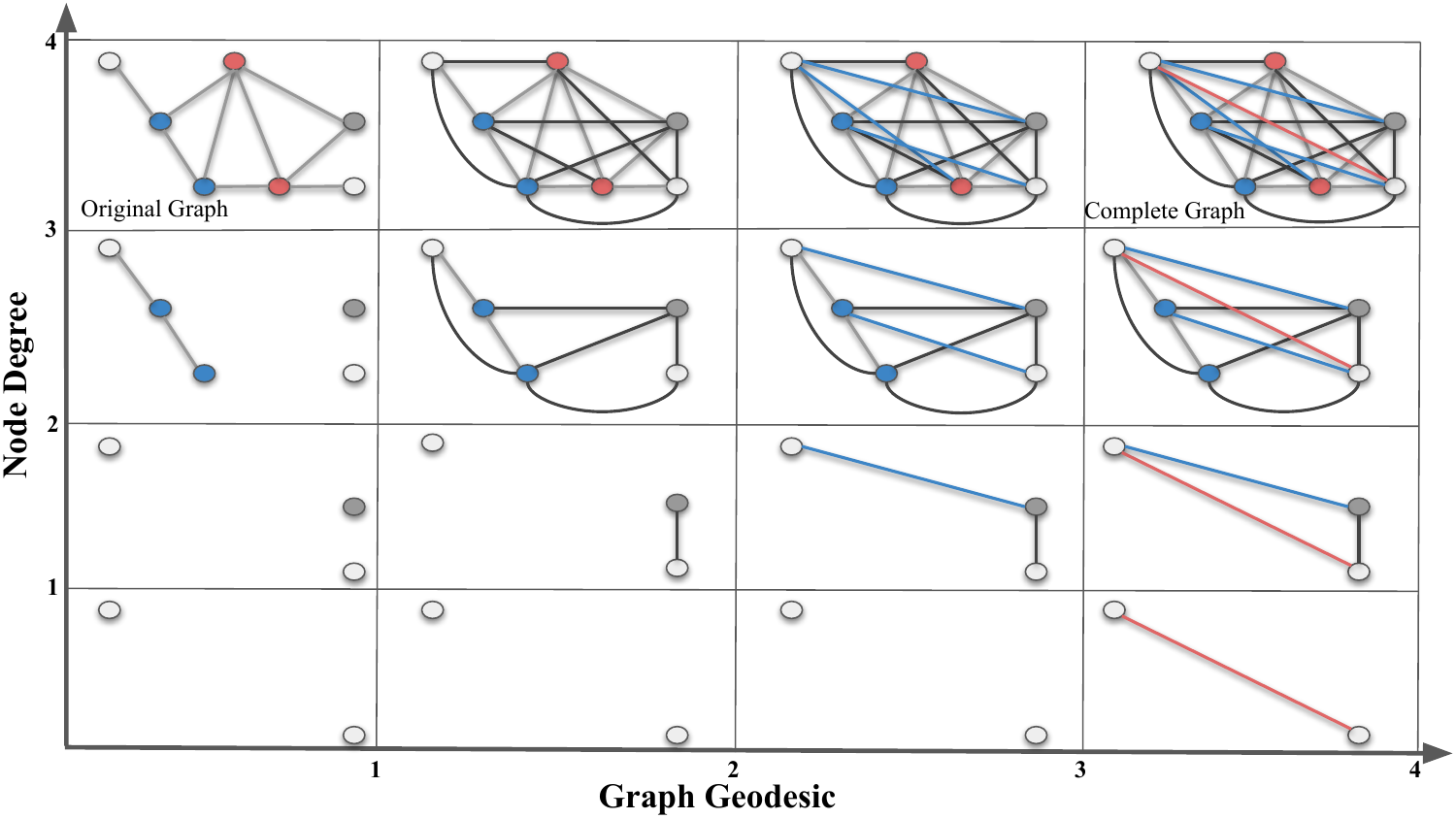}
\caption{\textbf{Vietoris-Rips (VR) Simplicial Filtrations.} \scriptsize In this illustration, a bifiltration is provided that combines a sublevel (vertical) and a Vietoris-Rips (VR) filtration (horizontal) of a simple graph $\G$ (top box in the first column). The vertical direction employs a sublevel filtration based on node degree with thresholds of $1,2,3$ and $4$. The horizontal direction employs a VR-filtration based on graph distance (geodesic length). The first column displays gray edges between nodes with a graph distance of $1$. The second column shows black edges between nodes with a graph distance of less than or equal to $2$. The third column displays blue edges between nodes with a graph distance of $3$, and the last column shows red edges between nodes with a graph distance of $4$.}
\label{fig:VR_Filtrations}
\end{figure}

Before constructing VR simplicial complexes, we compute the distances between each node in graph $\G$, i.e., $d(v_r,v_s)=d_{rs}$ is the length of the shortest path from $v_r$ to $v_s$ where each edge has length $1$. Let $K=\max d_{rs}$. Then, for each $1\leq i_0\leq m$, define VR-filtration for the vertex set $\V_{i_0}$ with the distances $d(v_r,v_s)=d_{rs}$, i.e., $\Delta_{i_00} \subseteq \Delta_{i_01} \subseteq \ldots \subseteq \Delta_{i_0K}$ (See Figure \ref{fig:VR_Filtrations}).  This gives $m\times (K+1)$ simplicial complexes $\{\Delta_{ij}\}$ where $1\leq i\leq m$ and $0\leq j\leq K$. This is called the \textit{bipersistence module}. One can imagine increasing sequence of $\{\V_i\}$ as vertical direction, and induced VR-complexes $\{\Delta_{ij}\}$ as the horizontal direction. In our construction, we fix the slicing direction as the horizontal direction (VR-direction) in the bipersistence module, and obtain the persistence diagrams in these slices.

The toy example in Figure~\ref{fig:VR_Filtrations} shows a small graph $\G$ instead of a real compound to keep the exposition simple. Our sublevel filtration (vertical direction) comes from the degree function. Degree of a node is the number of edges incident to it. In the first column, we simply see the single sublevel filtration of $\G$ by the degree function. In each row, we develop VR-filtration of the subgraph by using the graph distances between the nodes. Here, graph distance between nodes means the length of the shortest path (geodesic) in the graph where each edge is taken as length $1$.  Then, in the second column, we add the edges for the nodes whose graph distance is equal to $2$. In the third column, we add the (blue) edges for the nodes whose graph distance is equal to $3$. Finally, in the last column, we add the (red) edges for the nodes whose graph distance is equal to $4$. By construction, all the graphs in the last column must be a complete graph as there is no more edge to add.

After getting the bifiltration, for each $1\leq i_0\leq m$, we obtain a single filtration  $\V_{i_0}=\Delta_{i_00} \subseteq \Delta_{i_01}\subseteq \ldots \subseteq \Delta_{i_0K}$ in horizontal direction. Each threshold level of VR filtration provides a persistence diagram $PD(\V_{i_0})$. Hence, we obtain $m$ persistence diagrams $\{PD(\V_i)\}$. Then we apply a vectorization, $\varphi$, to each persistence diagram and obtain $m$ row vectors of fixed size $r$, i.e. $\vec{\varphi}_i=\varphi(PD(\V_i))$. This generates a $2D$-vector $\M_\varphi$ (a matrix) of size $m\times (K+1)$.


\noindent {\bf Persistence Diagrams:} After VR filtration, we systematically keep track of the evolution of topological patterns in the sequence of simplicial complexes $\{\wh{\G}_i\}_{i=1}^N$. A $k$-dimensional topological feature (or $k$-hole) may represent connected components ($0$-dimension), loops ($1$-dimension) and cavities ($2$-dimension). For each $k$-dimensional topological feature $\sigma$, persistent homology records its first appearance in the filtration sequence, say $\wh{\G}_{b_\sigma}$, first disappearance in later complexes, $\wh{\G}_{d_\sigma}$ 
with a unique pair $(b_\sigma, d_\sigma)$, where $1\leq b_\sigma<d_\sigma\leq N$.
We call $b_\sigma$ \textit{the birth time} of $\sigma$ and $d_\sigma$ \textit{the death time} of $\sigma$ and $d_\sigma-b_\sigma$ \textit{the life span} (or persistence) of $\sigma$. 
Persistence diagram records all these birth and death times of the topological features. Let $0\leq k\leq D$ where $D$ is the highest dimension in the simplicial complex $\wh{\G}_N$. Then $k^{th}$ persistence diagram ${\rm{PD}_k}(\G)=\{(b_\sigma, d_\sigma) \mid \sigma\in H_k(\wh{\G}_i) \mbox{ for } b_\sigma\leq i<d_\sigma\}$. Here, $H_k(\wh{\G}_i)$ represents the $k^{th}$ \textit{homology group} of $\wh{\G}_i$ which keeps the information of the $k$-holes in the simplicial complex $\wh{\G}_i$. We use $0$ and $1$ dimensional homology features, i.e., $PD_0(\G)$ and $PD_1(\G)$ in our implementation.

\noindent {\bf Vectorizations (Fingerprinting):} 
While PH extracts hidden shape patterns from data as persistence diagrams (PD), PDs being collection of points in $\RR^2$ by itself are not practical for statistical and machine learning (ML) purposes. Instead, the common techniques are by faithfully representing PDs as kernels~\citep{kriege2020survey} or vectorizations~\citep{hensel2021survey}. One can consider this step as converting PDs into a useful format to be used in ML process as fingerprints of the data. We use Betti curve vectorization~\citep{chung2022persistence} to transform Persistent Homology (PH) information represented as Persistent Diagrams (PDs) into a feature vector.

\section{Multiparameter Persistence (MP) Fingerprints} \label{sec:generalMP}

\subsection{Stability of MP Fingerprints} \label{sec:stability}

\noindent {\em Stability of Single Persistence Vectorizations:} 
A specific persistence diagram vectorization, denoted as $\varphi$, can be thought of as a mapping from the space of persistence diagrams to the space of functions. The concept of stability refers to the smoothness of this transformation. Essentially, it assesses whether a slight perturbation in the persistence diagram results in a significant change in the vectorization. To make this assessment meaningful, it is necessary to establish a metric in the space of persistence diagrams that defines what constitutes a ``slight perturbation''. The most commonly used metric for this purpose is the Wasserstein distance, also known as the matching distance, which is defined as follows.

Let $PD(\X^+)$ and $PD(\X^-)$ be persistence diagrams two datasets $\X^+$ and $\X^-$ (We omit the dimensions in PDs).  Let $PD(\X^+)=\{q_j^+\}\cup \Delta^+$ and  $PD(\X^-)=\{q_l^-\}\cup \Delta^-$ where $\Delta^\pm$ represents the diagonal (representing trivial cycles) with infinite multiplicity. Here, $q_j^+=(b^+_j,d_j^+)\in PD(\X^+)$ represents the birth and death times of a topological feature $\sigma_j$ in $\X^+$. Let $\phi:PD(\X^+)\to PD(\X^-)$ represent a bijection (matching). With the existence of the diagonal $\Delta^\pm$ in both sides, we make sure the existence of these bijections even if the cardinalities $|\{q_j^+\}|$ and $|\{q_l^-\}|$ are different. 
\begin{definition} Let $PD(\X^\pm)$ be persistence diagrams of the datasets $\X^\pm$, and $\mathbf{M}=\{\phi\}$ represent the space of matchings as described above.
Then, the $p^{th}$ Wasserstein distance $\W_p$ defined as $$\W_p(PD(\X^+),PD(\X^-))= \min_{\phi\in\mathbf{M}}\biggl(\sum_j\|q_j^+-\phi(q_j^+)\|_\infty^p\biggr)^\frac{1}{p}, \quad p\in \mathbb{Z}^+.$$
\end{definition}


Now, let's define the stability of vectorizations. A vectorization can be viewed as a mapping from the space of persistence diagrams, $\mathbf{P}$, to the space of functions or vectors $\mathbf{Y}$, for example, $\Psi:\mathbf{P}\to \mathbf{Y}$. In particular, if $\Psi$ is the persistence landscape, then $\mathbf{Y}=\mathcal{C}([0,K],\mathbb{R})$ and if $\Psi$ is the Betti summary, then $\mathbf{Y}=\mathbb{R}^m$. The stability of the vectorization $\Psi$ refers to the continuity of $\Psi$ as a mapping. Let $\mathrm{d}(.,.)$ be a suitable metric on the space of vectorizations. The stability of $\Psi$ can then be defined as follows:

\begin{definition} Let $\Psi:\mathbf{P}\to \mathbf{Y}$ be a vectorization for single persistence diagrams. Let $\W_p, \mathrm{d}$ be the metrics on $\mathbf{P}$ and $\mathbf{Y}$ respectively as described above. Let $\psi^\pm=\Psi(PD(\X^\pm))\in \mathbf{Y}$. Then, $\Psi$ is called \textit{stable} if $$\mathrm{d}(\psi^+,\psi^-)\leq C\cdot \W_{p_\Psi}(PD(\X^+),PD(\X^-))$$  
\end{definition}
In this context, the constant $C>0$ is independent of $\X^\pm$. The stability inequality states that the changes in the vectorizations are limited by the changes in persistence diagrams. The proximity of two persistence diagrams is reflected in the proximity of their respective vectorizations. A vectorization $\varphi$ is referred to as \textit{stable} if it satisfies this stability inequality for a given $\mathrm{d}$ and $\W_p$~\citep{atienza2020stability}.

Now, we are ready to show the stability of Multiparameter Persistent Fingerprints. Consider two graphs, $\G^+=(\V^+,\EE^+)$ and $\G^-=(\V^-,\EE^-)$. A stable SP vectorization is represented by $\varphi$, and it satisfies the stability equation,
\begin{equation}\label{eqn1}
\mathrm{d}(\varphi(PD(\G^+)),\varphi(PD(\G^-)))\leq C_\varphi\cdot \W_{p_\varphi}(PD(\G^+),PD(\G^-))
\end{equation}
for some $1\leq p_\varphi\leq \infty$. Here, $\varphi(\G^\pm)$ represent the corresponding vectorizations for $PD(\G^\pm)$ and $\W_p$ represents Wasserstein-$p$ distance as defined in Definition \ref{sec:stability}. 

Now,  let $f:\V^\pm\to \RR$ be a filtering function with threshold set $\{\alpha_i\}_{i=1}^m$. Then, define the sublevel vertex sets $\V^\pm_i=\{v_r\in\V^\pm\mid f(v_r)\leq \alpha_i\}$. For each $\V_i^\pm$, construct the induced VR-filtration  $\Delta^\pm_{i0}\subset\Delta^\pm_{i1}\subset \dots\subset\Delta^\pm_{iK}$ as before. For each $1\leq i_0\leq m$, we will have persistence diagram $PD(\V^\pm_{i_0})$ of the filtration $\{\Delta^\pm_{i_0k}\}$.

The induced matching distance between multiple persistence diagrams is defined as follows,
{\small 
\begin{equation}\label{eqn2}
 \mathbf{D}_{p,f}(\G^+,\G^-)\\=\sum_{i=1}^m\W_p(PD(\V^+_i), PD(\V^-_i)).
 \end{equation}  }
 
Now, we define the distance between induced MP Fingerprints as,
\begin{equation}\label{eqn3}
\mathfrak{D}_f(\M_\varphi(\G^+),\M_\varphi(\G^-))=\sum_{i=1}^m \mathrm{d}(\varphi(PD(\V^+_i)),\varphi(PD(\V^-_i)))
\end{equation}



\begin{theorem}
Let $\varphi$ be a stable SP vectorization. Then, the induced MP Fingerprint $\M_\varphi$ is also stable, i.e., with the notation above, there exists $\wh{C}_\varphi>0$ such that for any pair of graphs $\G^+$ and $\G^-$, we have the following inequality.
$$\mathfrak{D}(\M_\varphi(\G^+),\M_\varphi(\G^-))\leq \wh{C}_\varphi\cdot \mathbf{D}_{p_\varphi}(\{PD(\G^+)\},\{PD(\G^-)\})$$
\end{theorem}

\noindent {\em Proof:} 
As $\varphi$ is a stable SP vectorization, by Equation \ref{eqn1}, for any $1\leq i\leq m$, we have $\mathrm{d}(\varphi(PD(\V_i^+)),\varphi(PD(\V_i^+)))\leq C_\varphi\cdot \W_{p_\varphi}(PD(\V_i^+),PD(\V_i^-))$ for some $C_\varphi>0$ , where 
$\W_{p_\varphi}$ is Wasserstein-$p$ distance.
Notice that the constant $C_\varphi>0$ is independent of $i$. Hence, 
\begin{eqnarray*}
\mathfrak{D}(\M_\varphi(\G^+),\M_\varphi(\G^-)) & \quad  =  & \sum_{i=1}^m \mathrm{d}(\varphi(PD(\V_i^+)),\varphi(PD(\V_i^-))) \quad \quad \quad \quad  \quad \quad \\
\; & \quad \leq    & \sum_{i=1}^m C_\varphi\cdot \W_{p_\varphi}(PD(\V_i^+),PD(\V_i^-)) \quad \\
\; & \quad =  &  C_\varphi \sum_{i=1}^m \W_{p_\varphi}(PD(\V_i^+),PD(\V_i^-)) \quad \quad \\
\; & \quad =   & C_\varphi\cdot \mathbf{D}_{p_\varphi}(\G^+,\G^-) \ \quad 
\end{eqnarray*}

where the first and last equalities are due to Equation~\ref{eqn2} and Equation~\ref{eqn3}, while the inequality follows from Equation~\ref{eqn1} which is true for any $i$.

\subsection{Computational Complexity of MP Fingerprints}
\label{subsection: computation complexity}
The computational complexity (CC) of the MP Fingerprint $\mathbf{M}_\psi^d$ depends on the vectorization technique $\psi$ used and the number of filtering functions $d$. For a single persistence diagram $PD_k$, CC is $\mathcal{O}(\mathcal{N}^3)$, where $\mathcal{N}$ is the number of $k$-simplices~\citep{otter2017roadmap}. If $r$ is the resolution size of the multipersistence grid, then $CC(\mathbf{M}_\psi^d)=\mathcal{O}(r^d\cdot \mathcal{N}^3 \cdot C_\psi(m))$ where $C_\psi(m)$ is CC for $\psi$ and $m$ is the number of barcodes in $PD_k$,  e.g., if $\psi$ is Persistence Landscape, then $C_\psi(m)=m^2$ \citep{Bubenik:2015} and hence CC for MP Landscape with three filtering functions ($d=3$) is $\mathcal{O}(r^3\cdot \mathcal{N}^3\cdot m^2)$. Alternatively, for MP Betti summaries, the computation of persistence diagrams is not required. Instead, the rank of homology groups in the MP module must be determined. As a result, the computational complexity for MP Betti summary is significantly reduced by utilizing minimal representations~\citep{lesnick2019computing, kerber2021fast}. The feature extraction process is parallelized across the eight cores of an Intel Core i7 CPU (equipped with 100GB of RAM) through the use of multiprocessing. Further evaluation of the computation time for MP fingerprint extraction from datasets can be found in Table~\ref{modality_results}. In comparison to graph-based models that encode compounds by discovering common molecular fragments (known as motifs)~\citep{jin2020hierarchical}, all ToDD models necessitate fewer computational resources during their training phase.

\section{Experiments}
\subsection{Datasets \& Baselines}
\noindent {\bf Lipophilicity\footnote{\url{ https://deepchemdata.s3-us-west-1.amazonaws.com/datasets/Lipophilicity.csv}:}}  is a crucial characteristic of drug molecules that impacts both permeability through membranes and solubility. The dataset, sourced from the ChEMBL database, contains experimental results for the octanol/water distribution coefficient (logD at pH $7.4$) of $4200$ compounds.

\noindent {\bf FreeSolv\footnote{\url{https://deepchemdata.s3.us-west-1.amazonaws.com/datasets/freesolv.csv.gz}:}} is a compilation of both calculated and experimentally determined hydration free energies for $642$ small molecules in water.

\noindent {\bf ESOL\footnote{\url{https://deepchemdata.s3-us-west-1.amazonaws.com/datasets/delaney-processed.csv}}} is a collection of $1128$ chemical compounds and their corresponding water solubility values.

\begin{wraptable}{R}{8cm}
\centering
\caption{\footnotesize Comparison of RMSE performance between baseline models and ToDD on molecular property prediction task using MoleculeNet~\citep{wu2018moleculenet}, a large scale benchmark for molecular machine learning. Best score is highlighted in bold, and the best baseline is underlined.}
\label{baseline_comparison}
\setlength\tabcolsep{7 pt}
\scriptsize

\begin{tabular}{lc|c|c}
\toprule
\textbf{Model} & \textbf{Lipophilicity} & \textbf{FreeSolv} & \textbf{ESOL} \\
\midrule
Morgan (radius=4) & 0.817$\pm$0.045 & \underline{0.478$\pm$0.033} &  1.255$\pm$0.051 \\
Weave & 0.813$\pm$0.042 & 2.398$\pm$0.250 & 1.158$\pm$0.055  \\ 
SchNet & 0.909$\pm$0.098 & 3.215$\pm$0.755 & 1.045$\pm$0.064 \\
D-MPNN & 0.646$\pm$0.041 & 1.010 $\pm$ 0.064 & 0.980$\pm$0.258 \\
MGCN & 1.113$\pm$0.041 & 3.349$\pm$0.097 & 1.266$\pm$0.147 \\ 
Node-MPN & 0.672$\pm$0.051 & 2.185$\pm$0.952 & 1.167$\pm$0.430 \\ 
Edge-MPN & 0.653$\pm$0.046 & 2.177$\pm$0.914 & 0.980$\pm$0.258 \\
MV-GNN & \textbf{\underline{0.599$\pm$0.016}} & 1.840$\pm$0.019 & \underline{0.805$\pm$0.036} \\
GRAPHCL & 0.714$\pm$0.011 & 3.744$\pm$0.292 & 0.959$\pm$0.047 \\
3D Infomax & 0.695$\pm$0.012 & 2.337$\pm$0.227 & 0.894$\pm$0.028 \\
\toprule
\textbf{ToDD} & 0.738$\pm$0.025 & \textbf{0.354$\pm$0.053} & \textbf{0.612$\pm$0.083}  \\
\bottomrule
\end{tabular}
\end{wraptable}

We thoroughly evaluate the performance of our methods against the $11$ state-of-the-art baselines: 
Weave~\citep{kearnes2016molecular}, SchNet~\citep{schutt2017schnet}, 
Node-MPN~\citep{gilmer2017neural},
Edge-MPN~\citep{yang2019analyzing}
D-MPNN~\citep{yang2019analyzing},
MGCN~\citep{lu2019molecular}, 
OT-GNN~\citep{becigneul2020optimal},
MV-GNN~\citep{ma2020multi},
StructGNN~\citep{lukashina2020lipophilicity},
GRAPHCL~\citep{you2020graph} and
3D Infomax~\citep{stark20223d} on benchmark datasets.

\subsection{Experimental Results}
We train a Gradient Boosting Regression Tree (GBRT) model with $1000$ boosting stages using MP Fingerprints, namely ToDD~\citep{demir2022todd} as the input. The maximum depth of the trees was optimized to $6$ through tuning. The minimum required number of samples to split a node within the tree was 2. The Friedman loss function and a learning rate of $0.1$ were utilized for model optimization. The effectiveness of the proposed method was rigorously evaluated through a $5$-fold cross-validation, demonstrating its competitiveness against state-of-the-art GNN models as shown in Table~\ref{baseline_comparison}. ToDD is significantly better than the best MoleculeNet models on FreeSolv and ESOL, and is not substantially different on Lipophilicity. These findings suggest that ToDD surpasses the best MoleculeNet models while avoiding the need for training large-scale GNNs or generating $3D$ conformations. Furthermore, both ToDD and Morgan fingerprints exhibit a marked improvement over all GNN baselines on FreeSolv. However, it is important to note, that FreeSolv is a small scale dataset and the performance of GNNs can be impacted due to the limited number of training samples. Hypothetically, ToDD can demonstrate a significant advantage in leveraging domain knowledge to enhance property prediction scores with relative ease. For instance, it is a well-known observation that introducing nonpolar groups, such as methyl groups, as new substituents into a molecule can increase its lipophilicity. ToDD can effectively leverage bond polarity as an additional parameter to extract MP fingerprints, thereby integrating vital domain information for improved performance in lipophilicity prediction. 


\begin{table*}[htbp]
\centering
\caption{\footnotesize RMSE scores for single parameter and multiparameter PH fingerprints and clock time performance to extract Vietoris-Rips persistent homology features.} \label{modality_results}
\setlength\tabcolsep{3 pt}
\scriptsize

\begin{tabular}{lccccccccc}
\toprule
& \multicolumn{2}{c}{\textbf{Atomic Mass}}
& \multicolumn{2}{c}{\textbf{Partial Charge}}
& \multicolumn{2}{c}{\textbf{Bond Type}}
& \multicolumn{2}{c}{\textbf{All Parameters}}
\cr  
\cmidrule(lr){2-3} \cmidrule(lr){4-5} \cmidrule(lr){6-7} \cmidrule(lr){8-9}
\textbf{Dataset} & \textbf{RMSE} &\textbf{Time} & \textbf{RMSE} &\textbf{Time} &\textbf{RMSE} &\textbf{Time} &\textbf{RMSE} &\textbf{Time}\\
\midrule
Lipophilicity &0.989$\pm$0.027 &55 sec &0.989$\pm$0.034 &42 sec &0.995$\pm$0.049 &17 sec &\textbf{0.738$\pm$0.025} &114 sec\\ 
FreeSolv &0.605$\pm$0.124 &7 sec &0.557$\pm$0.099 &7 sec &0.894$\pm$0.180 &2 sec &\textbf{0.354$\pm$0.053} &16 sec\\ 
ESOL &0.929$\pm$0.066 &11 sec &0.997$\pm$0.103 &11 sec &1.340$\pm$0.078 &4 sec &\textbf{0.612$\pm$0.083} &26 sec\\
\bottomrule
\end{tabular}
\end{table*}

\subsection{Ablation Studies}
We investigate the impact of incorporating additional information about the domain on the model's performance. Table~\ref{modality_results} shows the results of evaluating the performance of single parameter persistence using only atomic mass, partial charge, and bond type. Afterwards, the Betti vectorizations from all the modalities are combined. Our findings indicate that the significance of single parameter persistence varies across the parameters, but combining the topological fingerprints learned from each modality into a single representation leads to a significant improvement in the RMSE scores, due to the complementary nature of the information sources.


\section{Conclusion}
A major challenge in the field of molecular machine learning is the conversion of molecules into concise fixed-length vectors. To address this issue, ToDD employs multiparameter persistent homology to construct a hierarchical topological representation of molecules. Our results indicate that this method has produced promising outcomes in predicting bioactivity and molecular properties, with the potential for further improvement through the integration of additional chemical parameters.

\clearpage
\bibliography{iclr2023_conference}
\bibliographystyle{iclr2023_conference}


\end{document}